\documentclass[twocolumn,superscriptaddress,showpacs,aps,floatfix]{revtex4-2}
\setcitestyle{square}
\usepackage[utf8]{inputenc}
\usepackage{graphicx,url, microtype, graphicx, dcolumn, bm, dsfont, csquotes, dsfont, amsmath, amsbsy,amsfonts,appendix, booktabs, amssymb, soul}
\usepackage[dvipsnames]{xcolor}
\usepackage[colorlinks,linkcolor=blue,citecolor=blue]{hyperref}

\usepackage[utf8]{inputenc}
\usepackage[T1]{fontenc}
\usepackage[margin=1in]{geometry}
\usepackage[sort&compress]{natbib}
\usepackage[skins,theorems]{tcolorbox}
\usetikzlibrary{intersections, angles, fadings, through, positioning,arrows,shapes,automata,petri,positioning,calc}

\newcommand{\bs}[1]{{\boldsymbol{#1}}}

\newcommand{\nhat}{\hat{\bs{n}}}

\pagecolor{white}

\definecolor{ResourceGreen}{RGB}{13, 101, 40}
\definecolor{RochesterBlue}{RGB}{0, 59, 113}
\definecolor{SkyBlue}{RGB}{67, 103, 118}
\definecolor{ForagerPink}{RGB}{153,124,129}
\definecolor{ForagerPink2}{RGB}{204,189,192}

\mathchardef\mhyphen="2D
\bibstyle{aps}

\begin{document}
\title{Semantic information in a model of resource gathering agents}
\author{Damian R Sowinski}
\email{Damian.Sowinski@rochester.edu}
\author{Jonathan Carroll-Nellenback}
\author{Robert N Markwick}
\affiliation{Department of Physics and Astronomy, University of Rochester}
\author{Jordi Pi{\~n}ero}
\affiliation{Complex Systems Lab, Pompeu Fabra University}
\author{Marcelo Gleiser}
\affiliation{Department of Physics and Astronomy, Dartmouth College}
\author{Artemy Kolchinsky}
\affiliation{Complex Systems Lab, Pompeu Fabra University}
\affiliation{Universal Biology Institute, The University of
 Tokyo, Japan}
\author{Gourab Ghoshal}
\email{gghoshal@pas.rochester.edu}
\affiliation{Department of Physics and Astronomy, University of Rochester}
\author{Adam Frank}
\email{afrank@pas.rochester.edu}
\affiliation{Department of Physics and Astronomy, University of Rochester}

\begin{abstract}
We explore the application of a new theory of \emph{Semantic Information} to the well-motivated problem resource foraging.  
Semantic information is defined as the subset of correlations, which is here measured via the transfer entropy, between agent $A$ and environment $E$ that is necessary for the agent to maintain its viability $V$.  
Viability, in turn, is endogenously defined as opposed to the use of exogenous quantities like utility functions.  
In our model, the forager's movements are determined by its ability to measure, via a sensor, the presence of an individual unit of resource, while the viability function is its expected lifetime. 
Through "interventions"--- i.e. scrambling the correlations between agent and environment via noising the sensor---we demonstrate the presence of a critical value of the noise parameter, $\eta_c$, above which the forager's expected lifetime is dramatically reduced. On the other hand, for $\eta < \eta_c$ there is little-to-no effect on its ability to survive. 
We refer to this boundary as the {\it semantic threshold}, quantifying the subset of agent-environment correlations that the agent actually needs to maintain its desired state of staying alive. 
Each bit of information affects the agent's ability to persist both above and below the semantic threshold. Modeling the viability curve and its semantic threshold via forager/environment parameters, we show how the correlations are instantiated. 
{Our work represents the first successful application of semantic information to a well-known agent-based model of biological and ecological interest. 
Additionally, we demonstrate that the concept of semantic thresholds may prove useful for understanding the role information plays in allowing systems to become autonomous agents.} 
\end{abstract}

\maketitle

\section{Introduction}
\label{sec: Introduction}
Questions about the role of information in the physics of life extend as far back as Schr\"odinger's seminal 1944 work ``What is Life'' ~\cite{schrodinger1944life}. 
Four years later, Shannon published his seminal work on  information theory \cite{shannon1948mathematical}, shortly followed by the discovery that DNA serves as a code for living organisms \cite{watson1953molecular}. 
These developments have led to a deep interest in  the relationship between information, physics, and biology~\cite{hopfield1994physics}. 
Since then, the applications of information theory to biology have grown exponentially~\cite{Smith_2007,mattingly2021escherichia,kuppers1990information,adami2004information,uda2020application,gohari2016information,rhee2012application,tkavcik2016information,donaldson2010fitness,rivoire2011value,hazen2007functional}, allowing researchers to unpack the ways organisms store and process data about the environment and their own internal states~\cite{sowinski2022consensus, oh2020towards, bowen2022visual}.

One difficulty with applications of Shannon's information theory to biological systems is its ``syntactic'' nature.
That is, the kinds of extant measures employed in information theory capture statistical correlations between systems without any consideration of the relevance or meaning of those correlations.
Living systems, however, act as agents for whom information is intrinsically meaningful in the most basic sense; that is, whether it can be useful for its {\it self-production} and {\it self-maintenance}~\cite{Schlosser_1998,Mossio_2009}.
Life, as a driven, non-linear, and far-from-equilibrium system, is always in a precarious position and must gather information about the state of the environment and its internal state to endure~\cite{Fang_2020}.
Some of this information will be useful for this purpose and some will be irrelevant.
In this setting, relevance and meaning can be considered synonymous.  

Unlike the well-developed field of syntactic information theory, there exists no widely accepted or applied formal theory of semantic information (previous attempts at developing such a theory include Refs.~\cite{Polani_2001,Thompson_2009,nehaniv_meaningful_2003,barham1996dynamical,deacon2007shannon,corning2007control, gleiser2018we}).
A goal of a {precise} mathematical theory {of semantic information} would be to provide an operational definition useful for characterizing nonlinear far-from-equilibrium systems which can be identified as agents (e.g., organisms or robots).
Recently, Kolchinsky and Wolpert \cite{kolchinsky2018semantic} (henceforth KW18) developed an explicit formalism for semantic information based on the use of counterfactuals and a notion of {\it viability}.
Their formulation uses the state spaces and probability distributions for an agent $A$ and its environment $E$ to characterize the mutual information between the two, while the persistence of $A$ (its ability to maintain a desired state) is measured through a viability function $V$.
The concept of meaning here is thus taken in the most basic sense of being related to an agent's continued existence.
By running {\it intervened} versions of the system dynamics in which some fraction of the mutual information between agent and environment  is scrambled, a formal working definition of the semantic information was characterized in terms of the response of the viability function to such interventions.  
Importantly, the viability is determined by the inherent coupled dynamics of the system and the environment~\cite{kolchinsky2018semantic,rovelli_meaning_2018}, rather than through exogenous utility, cost, error, or loss functions (as is sometimes done when studying the value of information in statistics or in engineering applications~\cite{gould_risk_1974,stratonovich_theory_2020,shannon1959coding,hazen2007functional,sowinski2021poroelasticity}).

{The {\it fitness value of information} has been recently proposed in an ecological setting, investigating the role of information in species growth using population distributions as proxy for probability measures~\cite{usinowicz2023fitness}.
We note that while the KW18 paper outlines the formalism for semantic information, there is yet to be any application of their framework in a biological setting, or indeed in any realistic individual organism model.
As we will see one reason for this is the need to deal with both the formal and computational issues associated with the extensive state spaces associated with the method.
}

There are several classes of models that describe specific attributes of living systems such as synchronization~\cite{Acebron_2005, Garcia_2021}, pattern formation~\cite{Cooper_2013, Mimar_2019}, competition between species for resources~\cite{Cohen_1977}, stability of ecosystems~\cite{Rooney_2006}, simple models of metabolism~\cite{Xavier_2020} as well as synthetic cells~\cite{Craig_2014}. 
Each of these models are settings where one can test the theoretical framework of semantic information. 
{Indeed, the only work to date that has used the KW18 formalism for semantic information was in the context of synthetic cells ~\cite{ruzzante2023synthetic}. 
However, in that short study no extension of the formalism was attempted.}

{In this paper, we make the first attempt to test the formalism of semantic information in a model of central interest to biology and ecology: the two-dimensional forager.
Our approach is three-fold: First, we apply the semantic information framework to the well-motivated problem of resource foraging agents.
Second, we develop a detailed mathematical and numerical implementation, operationalizing the original KW18 formalism to explore what simplifying approximations lead to a clear connection between semantic information and the viability of the foraging agents. 
Finally, we demonstrate the efficacy of our approach in uncovering new insights on the general features of agent/environment dynamics.}

Section~\ref{sec: Toy Model} focuses on a class of models that jointly address the interrelation between exploration and resource-consumption.
In such \emph{forager models}~\cite{benichou2014depletion, bhat2017does, benichou2016role, bhat2022smart, benichou2011intermittent, benichou2005stochastic, benichou2006two, bartumeus2002optimizing,viswanathan2002levy,james2008optimizing,campos2021optimal, bartumeus2016foraging}, an agent navigates its environment (using various exploration strategies) in search of resources (food) that it then consumes to maintain its desired internal state (staying alive).
The resources are either at fixed locations along the spatial extent of the environment or are replenished at a constant rate in random locations.
These dynamics are described in detail in section \ref{sec: 2A}.
In such a setting, a natural choice for a viability function is the agent's lifetime, while the environment is simply the field of resources.
The correlations between the agent and its environment, which are contingent on the agent's sensorial capability, carry meaningful information about the location of the resources.
These correlations are then scrambled by tuning the fidelity of the sensor and then measuring its effect on the viability function, the quantitative details of which make up section \ref{sec: 2B}.

We {describe our simulations and} demonstrate {numerically} in section~\ref{sec: 3} the existence of a plateau in the viability, capturing the subset of correlations that has little effect on the agent's ability to stay alive.
Below this threshold the lifetime of the agent monotonically decays with increased scrambling of the mutual information.
{Using geometric arguments, we derive an analytical expression for the threshold, relating it directly to sensor resolution.
We introduce a new semantic information concept, the \emph{viability-per-bit}, that captures the degree to which each bit of the agent's information on its environment is relevant.
This quantity, not articulated in the original KW18 formalism, peaks at the boundary separating the plateau from the decaying region indicating the existence of a \emph{semantic threshold}. 
Information above the threshold has little-to-no effect on the agent's viability, while below it, each bit becomes crucial for the agent to stay alive.

On the plateau, we investigate the appearance of a viability lower bound for a wide range of model parameters, then derive its analytical expression. Analytical and numerical evidence supporting these results is agnostic to the particular choice of viability function or indeed the type of foraging strategy.
We end in section \ref{sec: Conclusion} with a discussion of the implications of our findings and possible future directions. Five appendices clarify technical aspects of our model and the approximations used.}

\section{Foraging in a Replenishing Environment}
\label{sec: Toy Model}
{
Consider a system $A \times E$, defined to be a foraging agent, $A$, exploring an environment, $E$. 
The states of both agent and environment are described by finite collections of dynamical degrees of freedom, $a=(a_1,a_2,\dots)\in A$ and $e=(e_1,e_2,\dots)\in E$.
Evolution in a given time interval is described by a trajectory $\{(a(t),e(t))|t\in[t_i,t_f]\}$.
A {\it model} is the dynamical rule, deterministic or stochastic, that constructs this trajectory; the subset of the model describing the agent is referred to as a {\it foraging strategy}. 

There are several foraging strategies that have been considered in the literature including diffusive~\cite{benichou2014depletion, bhat2017does, benichou2016role, bhat2022smart}, intermittent~\cite{benichou2011intermittent, benichou2005stochastic, benichou2006two}, L\'evy-flight~\cite{bartumeus2002optimizing,viswanathan2002levy} and ballistic trajectories~\cite{james2008optimizing,campos2021optimal, bartumeus2016foraging}. 
In Section \ref{sec: 2A} we define in detail a model of a replenishing environment containing an agent with a ballistic foraging strategy.
The choice of this particular strategy stems from its analytical tractability, enabling the derivation of the semantic threshold and viability bound of section \ref{sec: 3}, which are then verified numerically. Other foraging strategies are then considered, and shown numerically to have similar results.
}
\subsection{Definition of the model}
\label{sec: 2A}

{We begin with a description of our foraging agent, and the environment it navigates.}
The agent has a fixed metabolic rate $\mu$ powered by an ever-dwindling fuel reserve, and a sensor allowing it to detect resources within some finite circular range $R$.
Moving at constant speed $v$, the agent changes its direction $\hat{\bs{n}}$ only in response to detected resources. 
Any resource that falls within the collection radius $r\le R$ of the agent can be harvested, its energy content added to the fuel supply of the agent.
The state of the agent is described by the tuple $a=(s,\tau,\hat{\bs{n}},\bs{x})$.
$s\in[0,S]$ is the agent's stored fuel supply; when $s=0$ the agent is no longer functional, otherwise it is considered {\it alive}.
The parameter $\tau\in\{0,1\}$ is Boolean and describes if the agent has locked on to a target. 
The agent will not change its direction of motion---indicated by the unit vector $\hat{\bs{n}}$---if it has a target.
Consuming a resource resets $\tau\mapsto0$.
Finally, $\bs{x}\in [0,L]^2$ is the position of the agent in the environment, defined as a square of side length $L\gg R$.

The state of the environment is specified by the set of locations of a fluctuating number of resources, $e=\{\bs{y}_1,\bs{y}_2,\dots,\bs{y}_N\}$ with $\bs{y}_n\in [0,L]^2,\ \forall n$.
The available resources are renewable---there is a source of energy flux per unit area into the environment, $\Gamma$, resulting in the growth of fixed energy $\epsilon$ resources across it, which, in turn, decay at a rate $\gamma$. 
In the absence of an agent, the two processes lead to an equilibrium average resource density $n_\text{eq}=\Gamma/\epsilon\gamma$ with $\mathcal O(L^{-1}\sqrt{\Gamma /\epsilon\gamma})$ fluctuations, so that the average distance between resources is $\ell_{\! Re}=\sqrt{\epsilon\gamma/\Gamma}$ (See Appendix~\ref{sec:erd} for details of the derivation).

Large inhomogeneities in the resource density, such as those which are created by a foraging agent, are repopulated at a rate $\gamma$.
We work in the small back-reaction regime where forager changes to the equilibrium spacing are on the order $\delta\ell_{\! Re}/\ell_{\! Re}\sim \mathcal O(R^2/L^2)\ll 1$. The agent is then effectively uncorrelated with the environmental degrees of freedom, on distances larger than some $\mathcal O(1)$ multiple of $R$.
Table~\ref{table: model parameters} summarizes the parameters that define the model. 

\begin{figure*}[ht]
    \centering
    \includegraphics[width = \textwidth]{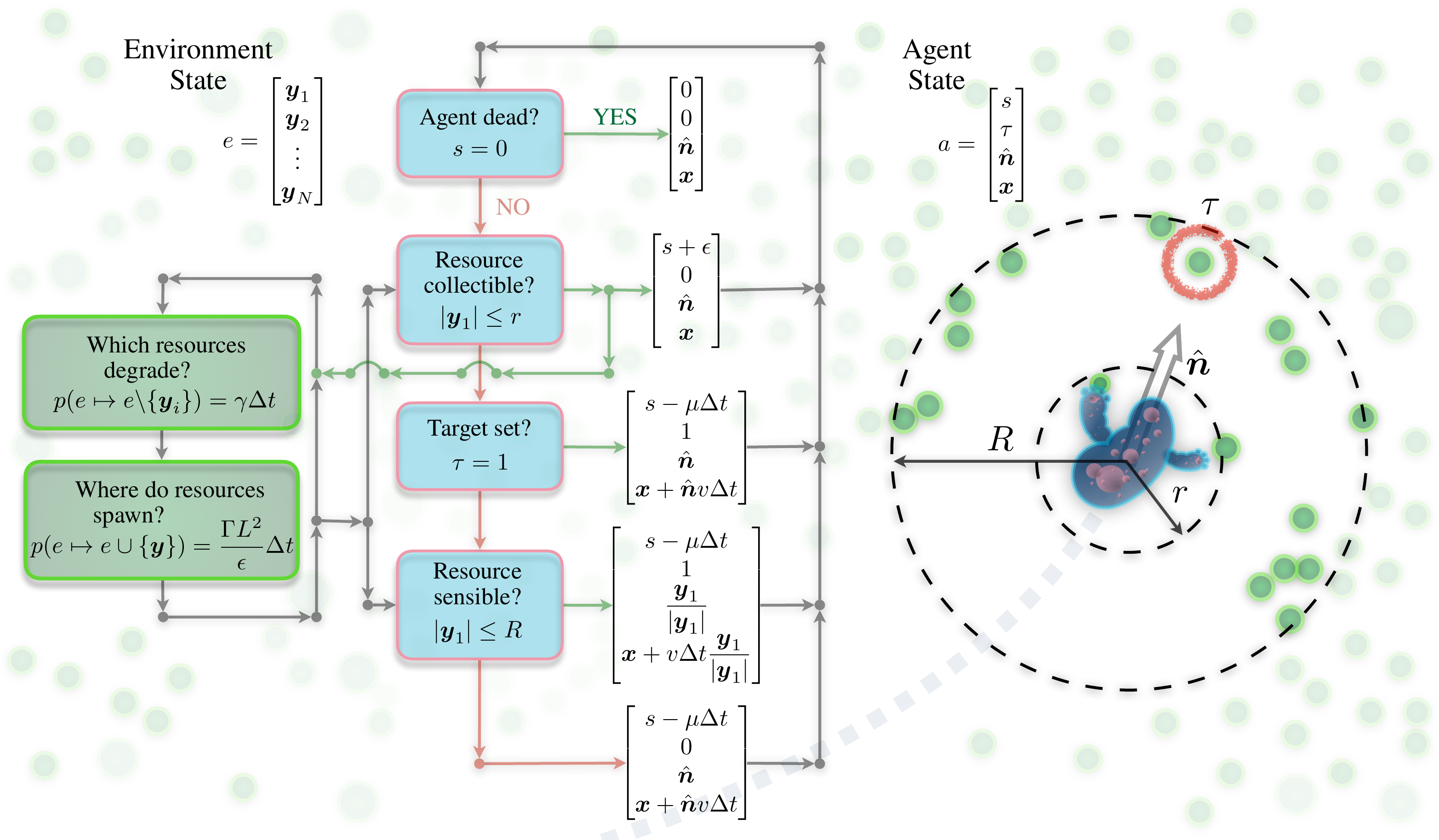}
    \caption{Overview of model with ballistic strategy. Agent moves at constant speed in a direction $\hat{\bs{n}}$, sensing at a distance $R$
  and collecting resources within radius $r$. The $2$d environment consists of decaying resources, randomly generated. Agent movement is powered by a metabolism that drains the agents energy reserves; collecting resources replenishes them. When a resource comes within sensor range, the agent targets (pink circle $\tau$), orients towards it, and then moves until the resource falls within collection range and is consumed.}
    \label{fig:forager_model}
\end{figure*}

\begin{table}[t!]
\caption{Model Parameters}
\centering 
\begin{tabular}{lcc}    \toprule
\emph{Parameter} & \emph{Symbol} & \emph{Units} \\\midrule
 Collection Radius    & $r$  & L  \\ 
 Detection Radius      & $R$  & L  \\
 Speed                & $v$  & LT$^{ \mhyphen 1}$\\
 Metabolic Rate       & $\mu$ & ML$^{2}$T$^{\mhyphen 3}$ \\
 Maximum Stored Energy & $S$ & ML$^{2}$T$^{\mhyphen 2}$ \\\midrule
 Resource Energy      & $\epsilon$ & ML$^{2}$T$^{\mhyphen 2}$ \\
 Resource Decay Rate  & $\gamma$ & T$^{\mhyphen 1}$ \\
 Energy Influx        & $\Gamma$ & MT$^{\mhyphen 3}$ \\
 \bottomrule
 \hline
\end{tabular}\label{table: model parameters}
\end{table}

As the agent moves through the environment, its collection diameter $2r$ acts as a cross-section, sweeping out an area for harvest per unit time proportional to its speed. 
Interesting regimes of the model happen when the speed falls within the following bounds:
\begin{equation}
    v_\star<v<v^\star\ \ \text{where} \ \ v_\star = \frac{\epsilon}{S}\frac{\mu\gamma}{2r\Gamma} \ \text{and} \ \ v^\star=\frac{\mu\gamma}{2r\Gamma}.
    \label{eq:vstar}
\end{equation}
For $ v < v_\star$, resources are spread so thin that the foraging area on a full tank has an expected harvest less than $1$, leading to agents with rather short lives. 
For $v > v^\star$, the harvest area fueled by a single resource has an expected harvest greater than $1$, the agent is coupled to an environment of abundance, and is effectively immortal.
In between, the agent's probability of survival decays exponentially at long lifetimes, while still forming significant correlations with the environment. This is the regime interesting for exploration. 
In a strong back-reaction setting these bounds are dynamic: agent harvesting increases the average spacing between resources thereby increasing both $v_\star$ and $v^\star$. Again, we will work in the small back-reaction limit, where the forager doesn't affect the environment.
Fig.~\ref{fig:forager_model} provides a schematic of the agent-centric dynamics.

\subsection{The viability function and interventions}
\label{sec: 2B}

While there are a number of ways of quantifying the viability of the agent, an obvious choice is the expected lifetime $V = \mathds{E}[T]$; here $\mathds{E}$ is an expectation value over an ensemble of agents, and $T$ is the time at which the agent first reaches the dead state ($s=0$) (as we will later show, our results are robust to other variants of the function).
This is easily measured by running an ensemble of agents as described in Fig.~\ref{fig:forager_model}, and characterizing the distribution of lifetimes. 
The viability of the agent depends crucially on it correctly setting its trajectory $\hat{\bs{n}}$ when a resource is targeted, which amounts to having a properly working sensor.
A broken sensor results in an agent moving in directions that do not correlate with the location of a resource, increasing the chance of starvation and thus death. 
{Importantly, when presenting our results below, we report viability values in units of $S/\mu$ (Maximum Stored Energy divided by Metabolic Rate, see Table~\ref{table: model parameters}). 
In these units, a viability value of 1 represents the expected lifetime of an agent that starts with a full tank and does not acquire any resources.}

To escape the vagaries of chance, the agent needs a sensor that correlates target direction with resource location. 
Let $\rho_{\!AE}(a,e)$ be the joint distribution over states of the agent and environment.
We use a standard notation where the subscript identifies the state space that the arguments are drawn from.
Without loss of generality, we assume that resources are indexed in order of distance to agent, with $\bs{y}_1$ being the closest.
To hone in on the specific correlation between sensor and environment, we next marginalize over all but the closest resource, $\bs{y}=\bs{y}_1$,  and condition on the agent being alive, leading to the simpler joint distribution $\rho_{AE}(\nhat, \tau,\bs{y}|s>0)$.
Rather than computing the correlation between all the agent and environmental degrees of freedom, we now have the simpler task of finding the correlations between the subsets $\{\nhat,\tau\}$ and $\{\bs{y}\}$.

Denote the living agent distribution $\rho_A = \rho_A(\nhat,\tau)$ and the environment distribution $\rho_E=\rho_E(\bs{y})$.
Correlations between agent and environment prevent a factorization of the joint distribution so that $\rho_{AE}\neq\rho_A\rho_E$.
(The distribution factorizes in the case of a non-functional agent, given that the lack of an interaction pathway rapidly suppresses correlations between the agent and environment.) 
While alive, interactions transfer information between the two, generating the correlations which are then responsible for prolonging the agent's viability.
Consequently, we examine the flow of information between the environment and the agent during a detection event.

Consider a resource detection and locking event during the time interval $\Delta t$, when a resource is detected at a distance $r<|\bs{y}|\le R$. 
The agent's orientation evolves from $\nhat\mapsto\nhat'$ while its target degree of freedom flips from $\tau=0\rightarrow \tau'=1$.
For an agent with a perfect sensor, the transition probability {between non-targeting and targeting agent states, $A\rightarrow A'$,} conditioned on the environment state, $e$, is
\begin{align}\label{eq: transition probability}
    \rho_{A\rightarrow A'|e}(\nhat',\tau'|\nhat,0,\bs{y}) = \delta_{\tau'}^1\delta\left(\theta(\nhat',\bs{y})\right),
\end{align}
where $\theta(\bs{a},\bs{b})$ is the angle between vectors $\bs{a}$ and $\bs{b}$, $\delta^{1}_{\tau'}$ is the Kronecker delta, and $\delta(\theta)$ is the Dirac delta function. 
By the rotational symmetry of the environment, we also have that  $\rho_{A\rightarrow A'}(\nhat',\tau'|\nhat,0) = \delta_{\tau'}^1(2\pi)^{\mhyphen 1}$, which is not equal to Eq.~\eqref{eq: transition probability} indicating that $\rho_{AE}\neq\rho_A\rho_E$.

To investigate the extent these correlations are necessary to ensure the agent's longevity, we compromise the agent's ability to find resources by adding noise to its sensor. 
This will impact the agent's viability function $V$. 
We thus add a small amount of noise $\eta> 0$ to the agent's ideal sensor, which affects the transition probability for detection defined in Eq.~\eqref{eq: transition probability} as $\rho_{A\rightarrow A'|e}\equiv \rho_{A\rightarrow A'|e}^{\eta=0}\mapsto \rho_{A\rightarrow A'|e}^\eta$. 
We measure the effect on the agent's viability as $\eta$ is increased to $1$, where $\rho_{A\rightarrow A'|e}^{\eta =1}\equiv\rho_{A\rightarrow A'}$. 
Adding noise changes the Dirac delta in Eq.~\eqref{eq: transition probability} into a uniform distribution of width $2\pi\eta$, $\delta\mapsto\delta_\eta$, where  
\begin{align*}
    \delta_\eta(\theta) =
    \begin{cases}
        \dfrac{1}{2\pi\eta} & \theta\in[\mhyphen \pi\eta,\pi\eta] \\
        0 & \text{otherwise}
    \end{cases}.
\end{align*}
As the noise parameter $\eta\rightarrow 0$, the perfect sensor is recovered, while $\eta\rightarrow 1$ gives the uniform distribution.
\begin{figure}[t!]
    \includegraphics[width = \columnwidth]{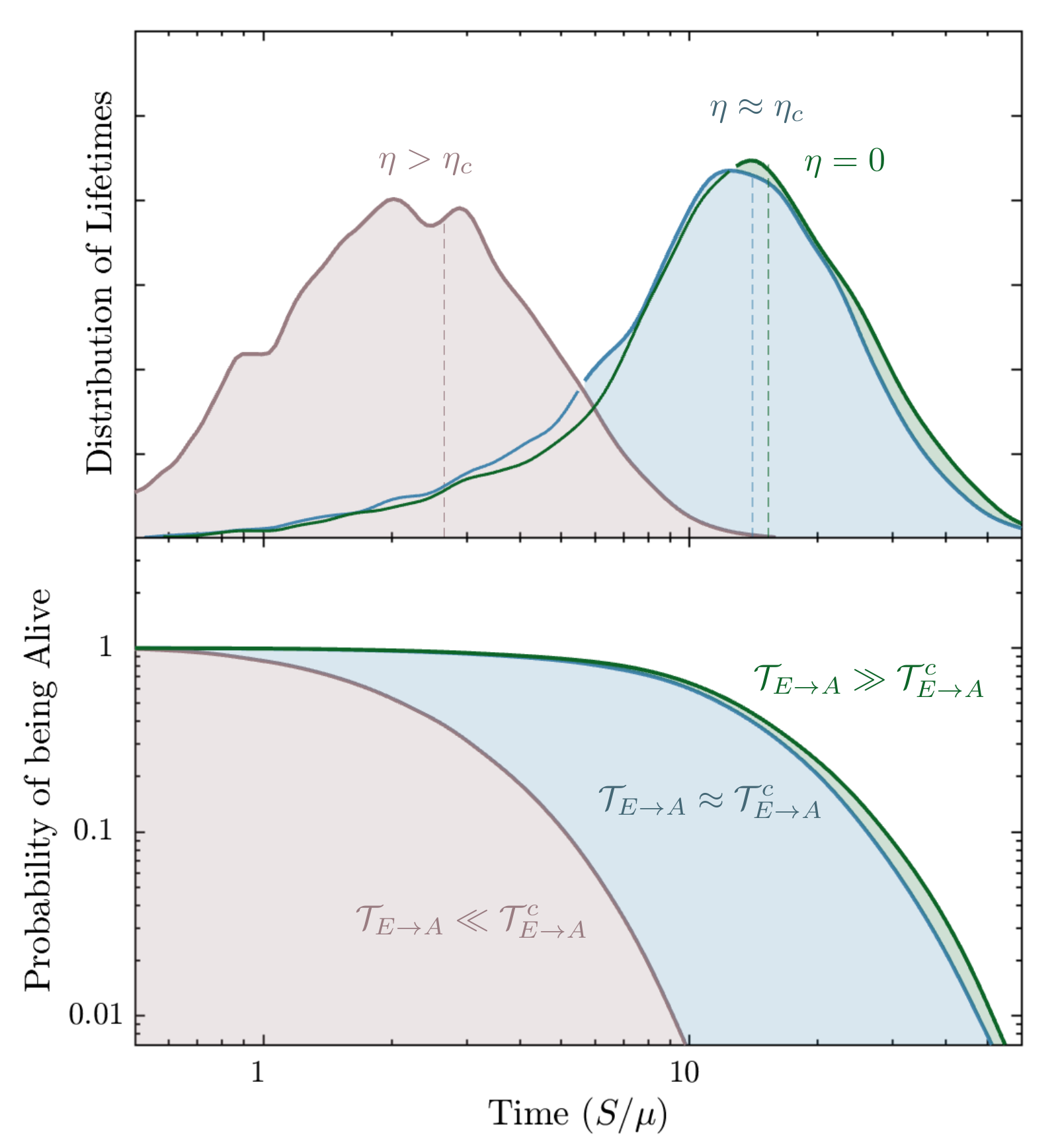}
    \caption{(Upper panel) The distribution of forager lifetimes extracted from the simulations. The green curve is for a noiseless (ideal) agent sensor ($\eta=0$). The nearly indistinguishable blue curve is from near critical scrambling ($\eta \approx \eta_c$), while the red curve is for above critical scrambling ($\eta>\eta_c$).  (Bottom panel) The corresponding distributions for the probability of being alive. Note the decay of the tails is faster than any power law.}
    \label{fig: Lifetimes}
\end{figure}

We approximate the information flowing from the environment to the agent during the detection event with the conditional transfer entropy (from the closest resource to the agent's direction, given the agent's state)~\cite{Schreiber_2000},
\begin{align} \label{eq: mutual information}
    \mathcal T_{E\rightarrow A}^\eta &= \mathds{E}_{\rho_{AE}}\left[\log_2 \frac{\rho_{A\rightarrow A'|e}}{\rho_{A\rightarrow A'}}\right]\nonumber\\
    &=\log_2\frac{1}{\eta}.
\end{align}
The information gathered during detection approaches $0$ as $\eta\rightarrow 1$, preventing the formation of correlations between agent and environment. 
It diverges as $\eta\rightarrow 0$, which reflects the infinite precision needed to specify a direction in space, as discussed in \cite{gleiser2018configurational}.

\section{Results}\label{sec: 3}
Returning to the notion of viability, we extend our definition to the noised sensor channel by defining $V = V_0\mapsto V_\eta =\mathds{E}_\eta[T]$, where the expectation is now taken over an ensemble of agents with noisy sensors. 
We interpret the addition of noise as a counterfactual intervention that scrambles the information transfer between an agent with a perfect sensor and its environment.
We extract $V_\eta$ by simulating {a single agent} within an equilibrated environment at half of full health and measuring how long {it survives}. 
This is done $10^4$ times for each of $200$ values of $\eta\in [0,1]$, the first $100$ equally spaced {in $[0,0.15)$ and the second hundred logarithmically spaced in $[0.15,1]$. 
Further details of the simulation procedure, including the range of values used for the parameters, is described in Appendix~\ref{app:sims}.
For animations of several instances of instances of agents with varying values of $\eta$, see the Supplemental Material~\cite{supplementary}.

\subsection{Viability and distribution of lifetimes}
In the upper panel of Fig.~\ref{fig: Lifetimes}, we plot the distribution of lifetimes for three different values of the noise parameter $\eta$. 
The green curve corresponds to the case of a perfect sensor (no scrambling, $\eta=0$) and sets the baseline distribution of lifetimes which is peaked near the expected lifetime (vertical dashed dark-green line) with a broadly decaying tail.
The red curve corresponds to a larger value of $\eta$ whereby one sees a much lower value for $\mathds{E}_\eta[T]$ once a critical value of noise, $\eta_c$ (blue curve) is surpassed. (Later in the section we derive this threshold analytically). In the bottom-panel, we show the corresponding distributions for the probability of an agent to stay alive, where one can see exponentially decaying tails.}

\begin{figure}[t!]
    \includegraphics[width = \columnwidth]{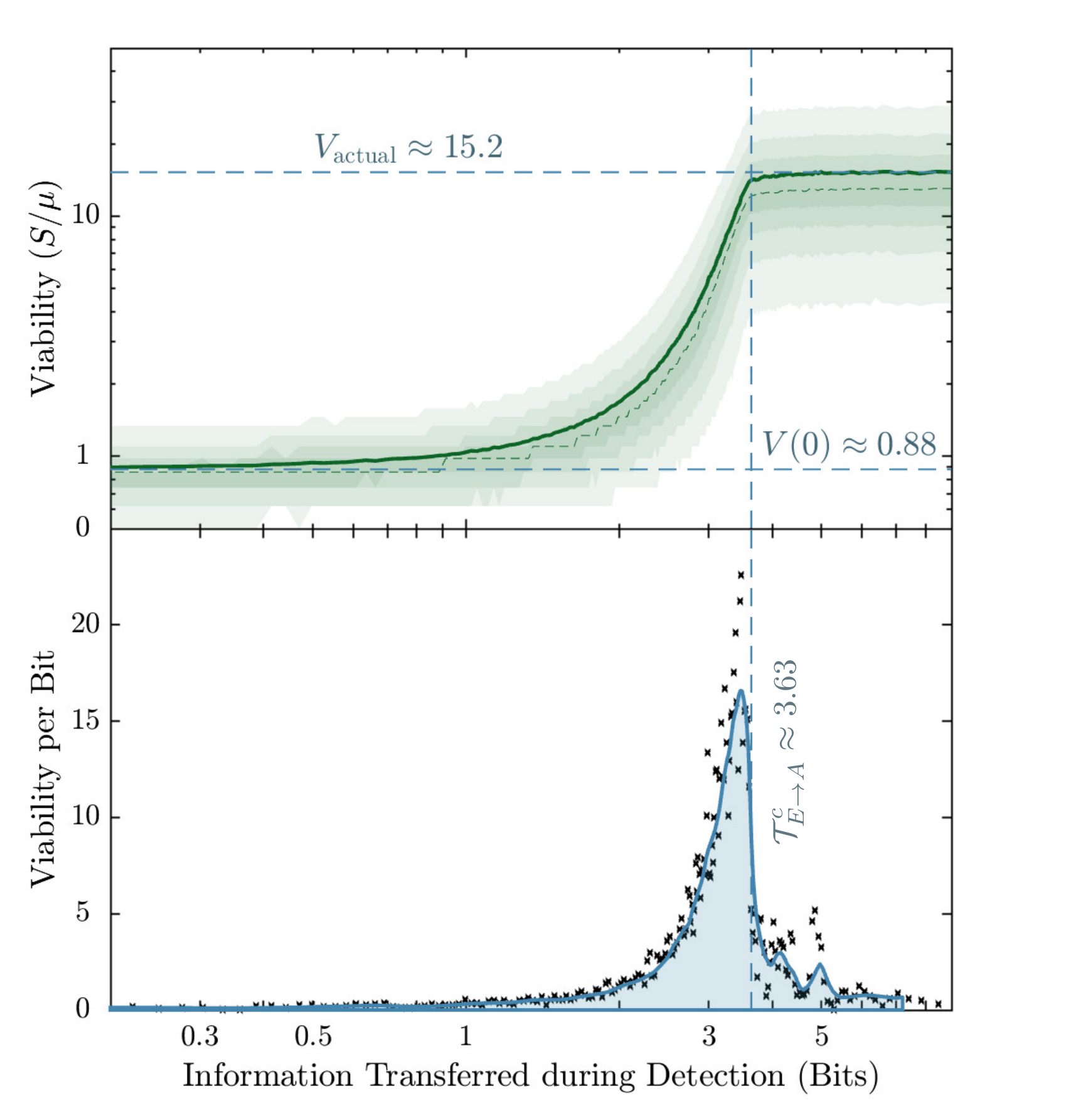}
    \caption{(Upper panel) The viability curve (thick green line) as a function of the transfer entropy, $\mathcal T_{E\rightarrow A}^\eta$. The actual viability $V_{\rm actual}$ is the expected lifetime with no scrambling. The light green line is the median, with the shaded regions representing intervals of $10\%$ above and below the median. 
    $V(0)$ indicates minimal viability achieved by an agent with a completely noisy sensor, see Eq.~\eqref{eq: broken sensor viability}. 
    (Bottom panel) The viability per bit. The dots are actual data, and the line is a smoothed and interpolated curve added to better visualize the semantic region. The vertical dashed line is the semantic threshold Eq.~\eqref{eq: semantic threshold}.}
    \label{fig: Viability}
\end{figure}

The actual viability $V_{\rm actual}$ is defined as the expected lifetime of an agent with noiseless sensor.
As anticipated in \cite{kolchinsky2018semantic}, the viability plateaus near the actual viability $V_{\rm actual}$ even as noise degrades the information flow during detection. However, once the sensor efficiency has been sufficiently degraded, the viability begins to drop dramatically, as seen in the top panel of Fig.~\ref{fig: Viability}.
This rapid decline in viability is interpreted as a semantic threshold---it is the minimal information acquired in a sensing event responsible for maintaining agents close to their actual viability. 
To better characterize this critical information, we introduce the differential measure of viability per bit (VpB $=dV/d\mathcal T$), as seen in the lower panel of Fig.~\ref{fig: Viability}.
The peak in VpB illustrates what an agent requires from its interactions with the environment.
Very noisy sensors provide little VpB, even with increased accuracy.
However, once a certain level of accuracy is achieved, the VpB grows dramatically and the sensor provides the agent with meaningful information about its environment, resulting in the agent's improved survivability.
Too much accuracy, however, is wasted, and the VpB drops again once the sensor goes above and beyond the constraints dictated by the physical nature of the agent within the environment. 

\subsection{Robustness to parameters and the viability bound}
The results in Fig.~\ref{fig: Viability} don't depend on using the expected lifetime as a viability function; any percentile of the lifetime distribution will do. 
The dashed green line shows the median of the distribution, while the {shaded green} regions represent jumps by $\pm10\%$ from the median. In all cases the qualitative shape of the curve and the semantic threshold is retained. This feature can be understood by considering how scrambling the sensor relates to the foraging efficacy of the agent.
Recall that the collection radius $r$ acts as an impact parameter, and that targets are set when they enter within the sensing radius $R$. 
For this reason, there is a critical noise parameter above which the agent will sometimes miss its target, and therefore achieve a lower viability. 
This critical value, and the corresponding semantic threshold, are
\begin{align}\label{eq: critical noise}
    \eta_c &= \frac{1}{\pi}\sin^{\mhyphen 1}\frac{r}{R}\\
\label{eq: semantic threshold}
    \mathcal T_{E\rightarrow A}^c &= \log_2\pi - \log_2\sin^{\mhyphen 1}\frac{r}{R}.
\end{align}
(See Appendix~\ref{sec:ST} for details.)
Interestingly, this expression does not depend on most parameters that characterize the agent, such as its speed and metabolic rate, only on its its geometry.
There is low VpB when the information transferred during a detection event is well above $\mathcal T_{E\rightarrow A}^c$, and high VpB as $\mathcal T_{E\rightarrow A}^\eta\rightarrow \mathcal T_{E\rightarrow A}^c$.

Above the semantic threshold ($\eta < \eta_c$), the minute drop in the viability plateau away from $V_\text{actual}$ also has a geometric origin.
Due to the circular shape of both the resource collection and sensing zones, a sensor with $\eta < \eta_c$ will force agents to travel farther ever so slightly to collect a targeted resource.
Assume an agent targets a resource a distance $y=|\bs{y}|\le R$ away, but the sensor is mistaken by an angle $\theta$ as to the direction of the resource.
Rather than having to travel $y-r$ to collect its target, the agent now has to travel $y\cos\theta - \sqrt{r^2-y^2\sin^2\theta}$.
Given an information transfer of $\mathcal T$ during detection, averaging over all angles we see that resources a distance $y$ from the agent are effectively a distance $(1+\lambda)y$ with 
\begin{align}
\label{eq: dilation0}
\lambda& = \frac{r}{y}\left(\!1\!-\!\frac{2^{\mathcal T}}{\pi}E\left(\frac{\pi}{2^{\mathcal T}}\Big|\frac{y^2}{r^2}\right)\!\right)\!-\!\left(\! 1\!-\!\frac{2^{\mathcal T}}{\pi}\sin \!\frac{\pi}{2^{\mathcal T}}\!\right) .
\end{align}
Here $E(\varphi|k^2)$ is the indefinite elliptic integral of the second kind with modulus $k=y/r$. (See Appendix~\ref{sec:vp} for details of the derivation.)
{We may consider two interesting limiting regimes. 
In the regime of accurate sensors, $\eta \approx 0$ and $\mathcal T\gg 1$, the dilation decays exponentially as
\begin{align}
    \label{eq:dilation1}
\lambda \sim \frac{\pi^2}{6}\frac{y-r}{r}4^{-\mathcal T}\,.
\end{align}
Another interesting regime to consider is when $y$ is very close to (but above) the Collection Radius $r$. In this case,
\begin{align}
\lambda \sim \frac{y-r}{r}\left(\frac{\tanh^{\mhyphen 1}\tan(\frac{\pi}{2}2^{\mhyphen \mathcal T})}{\frac{\pi}{2}2^{\mhyphen \mathcal T}}-1\right)\,,
\label{eq:dilation2}
\end{align}
up to first order in $y-r$.
As expected, the dilation grows linearly away from $r$, with the slope being controlled by the transfer entropy.}

The expression for the dilation factor $\lambda$ indicates that, above the semantic threshold ($\eta < \eta_c$), agents can be thought of as unscrambled agents making their way through a distorted world.
As the information acquisition from the environment is restored ($\mathcal T\rightarrow\infty$), the distortion decays exponentially as expected.
Of course, within the extraction radius  there is no dilation as the scrambled agent still has the reach $y=r$, even if it cannot judge properly distances $y> r$.
Maximal dilation $\lambda^{\max}$, occurs at $y=R$ in Eq.~\eqref{eq: dilation0}. This constitutes an upper bound, which enables the determination of a lower bound on the viability of a scrambled agent, given the actual viability of an unscrambled agent.

\begin{figure}[t!]
    \includegraphics[width = \columnwidth]{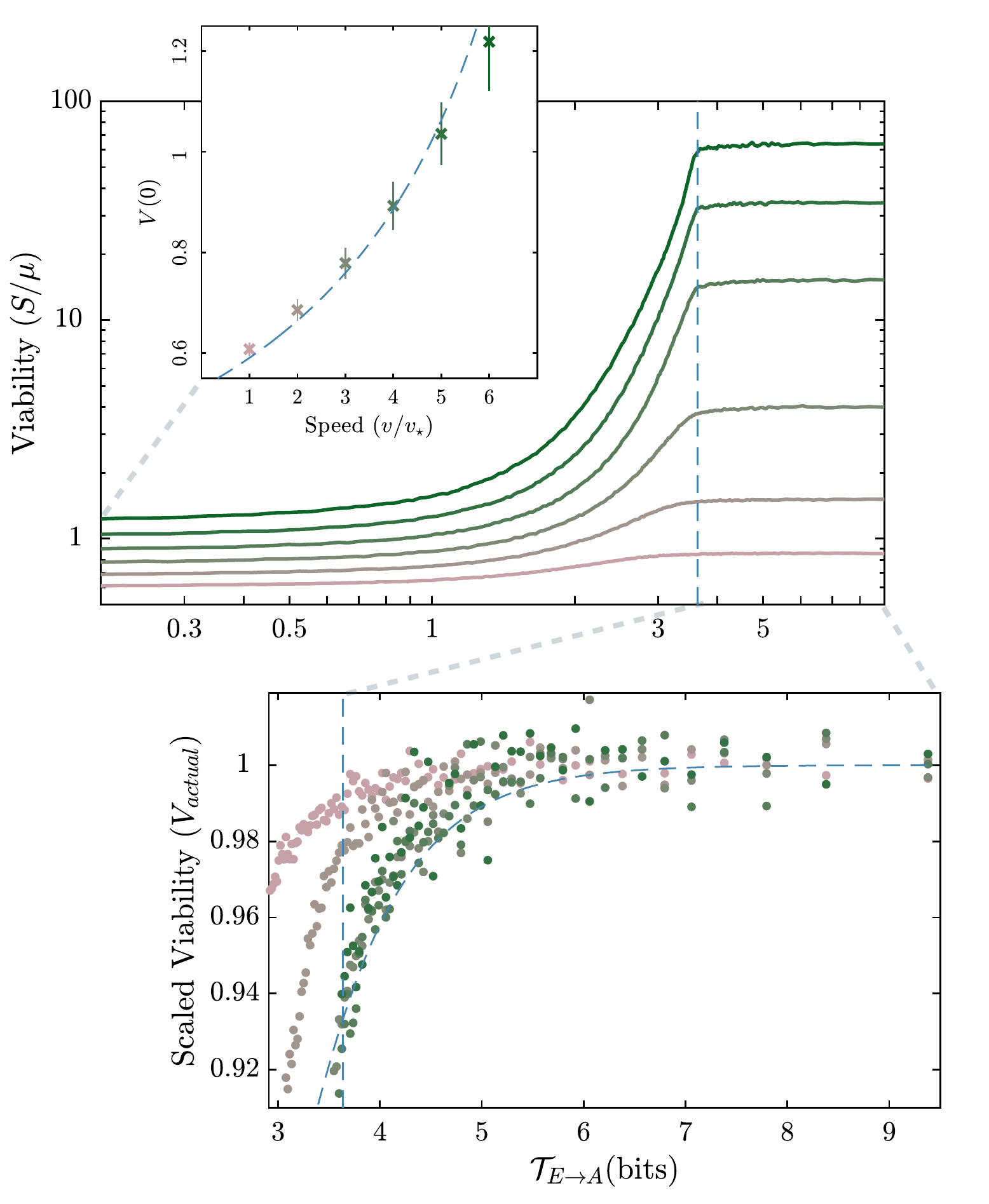}
    \caption{(Top panel) Viability curves for six velocities in the range $1 \leq v/v_{\star} \leq 6$ in integer increments. Here $v^\star=10v_\star$. The vertical dashed line is Eq.~\eqref{eq: semantic threshold}. 
    (Inset) Minimal viability as achieved by an agent with a completely noisy sensor, as $\mathcal T_{E\rightarrow A}\rightarrow 0$. Bars represent $5\sigma$ errors in the estimates. The dashed line represents Eq.~\eqref{eq: broken sensor viability}, showing good agreement with simulations. 
    (Bottom panel) The plateau region of each curve, rescaled for purposes of comparison. The dashed curve represents the viability bound Eq.~\eqref{eq: viability lower bound}.}
    \label{fig: Velocity}
\end{figure}

To compare the viability of scrambled and unscrambled agents, we scale down the length scale of the scrambled agent by $(1+\lambda^{\max})$ so that it matches the length scale of the unscrambled agent.
Unfortunately, this means the speeds of the two no longer match; to restore equality we must scale down the timescale by the same amount.
Numerical simulations indicate that effects of this rescaling does not affect the environmental variables. Focusing therefore only on the agent variables, the rescaled viability is
\begin{align}
V_\eta = \mathds{E}_\eta[T]
&=\mathds{E}_0[(1+\lambda^{\max})^{\mhyphen1}T]\nonumber \\
&=(1+\lambda^{\max})^{\mhyphen 1}V_\text{actual}.
\end{align}
Finally, expanding this expression for $\lambda^{\max}\ll 1$, above the semantic threshold, the viability as a function of the transfer entropy is bounded from below by
\begin{align}\label{eq: viability lower bound}
\frac{V(\mathcal T)}{V_\text{actual}} \gtrsim & \ \ 2-\frac{r}{R}+\frac{2^{\mathcal T-\mathcal T^c}}{\sin^{\mhyphen1}\frac{r}{R}}\nonumber\\
&\times\!\! \left[\frac{r}{R} E\left(\left.\frac{\sin^{\mhyphen1}\frac{r}{R}}{2^{\mathcal T-\mathcal T^c}}\right|\frac{R^2}{r^2}\right)\!-\!\sin\left(\frac{\sin^{\mhyphen1}\frac{r}{R}}{2^{\mathcal T-\mathcal T^c}}\right)\right],
\end{align}
where $\mathcal{T}^c$ is defined in Eq.~\eqref{eq: semantic threshold}.

It is instructive to check whether the results presented thus far are robust to varying the speed of the forager.  
As shown in Eq.~\eqref{eq:vstar}, the speed is naturally bounded in the limits $v_{\star} < v < v^{\star}$. 
In our simulations, we set $v^\star=10v_\star$; varying the velocity within this range, in the upper panel of Fig.~\ref{fig: Velocity}, we show the viability curves for six values of $1 \leq v/v_{\star} \leq 6$ in integer increments (colors represent different values of the ratio). 
The vertical dashed line is the analytically calculated semantic threshold (Eq.~\eqref{eq: semantic threshold}). 
As expected, while the height of the viability curve decreases with decreasing velocity (eventually flattening near the lower bound $v_{\star}$), the semantic threshold as well as the shape of each curve is robust to the change in velocities. 

In the bottom panel we zoom in to the plateau region for each curve, rescaling them for purposes of comparison. 
The viability bound calculated in Eq.~\eqref{eq: viability lower bound} is plotted as a dashed curve, while the semantic threshold is shown as a vertical dashed line. 
While the agreement with the viability bound improves with increasing velocity, Eq.~\eqref{eq: viability lower bound} bounds the plateau for all velocity ranges (up to statistical fluctuations).

One can estimate the minimal viability, $V(0)=\mathds{E}_{1}[T]$, as achieved by an agent with a completely noisy sensor. 
To do so, note  that resources encountered by such an agent follow a Poisson distribution with rate $2vr/\ell_{Re}^2$.
The total fuel processed by the agent will therefore be $s_0+\epsilon\mathds{E}_1[T]$, where $s_0$ is the initial stored fuel.
But this total fuel divided by the metabolic rate is precisely the expected lifetime; a little algebra reveals
\begin{equation}\label{eq: broken sensor viability}
    V(0) = \frac{s_0/\mu}{1-v/v^\star},   
\end{equation} 
where $v^\star$ is the upper velocity limit defined in Eq.~\eqref{eq:vstar}. 
As mentioned earlier, the agent approaches immortality when its velocity approaches this limit, even if it has a completely random sensor. 
In the inset panel of Fig.~\ref{fig: Velocity}, we plot Eq.~\eqref{eq: broken sensor viability} as a dashed line. 
Points indicate the minimal viability extracted from simulations as a function of velocity, showing excellent agreement with the theoretical prediction.

\begin{figure*}[ht]
    \includegraphics[width = 1.8\columnwidth]{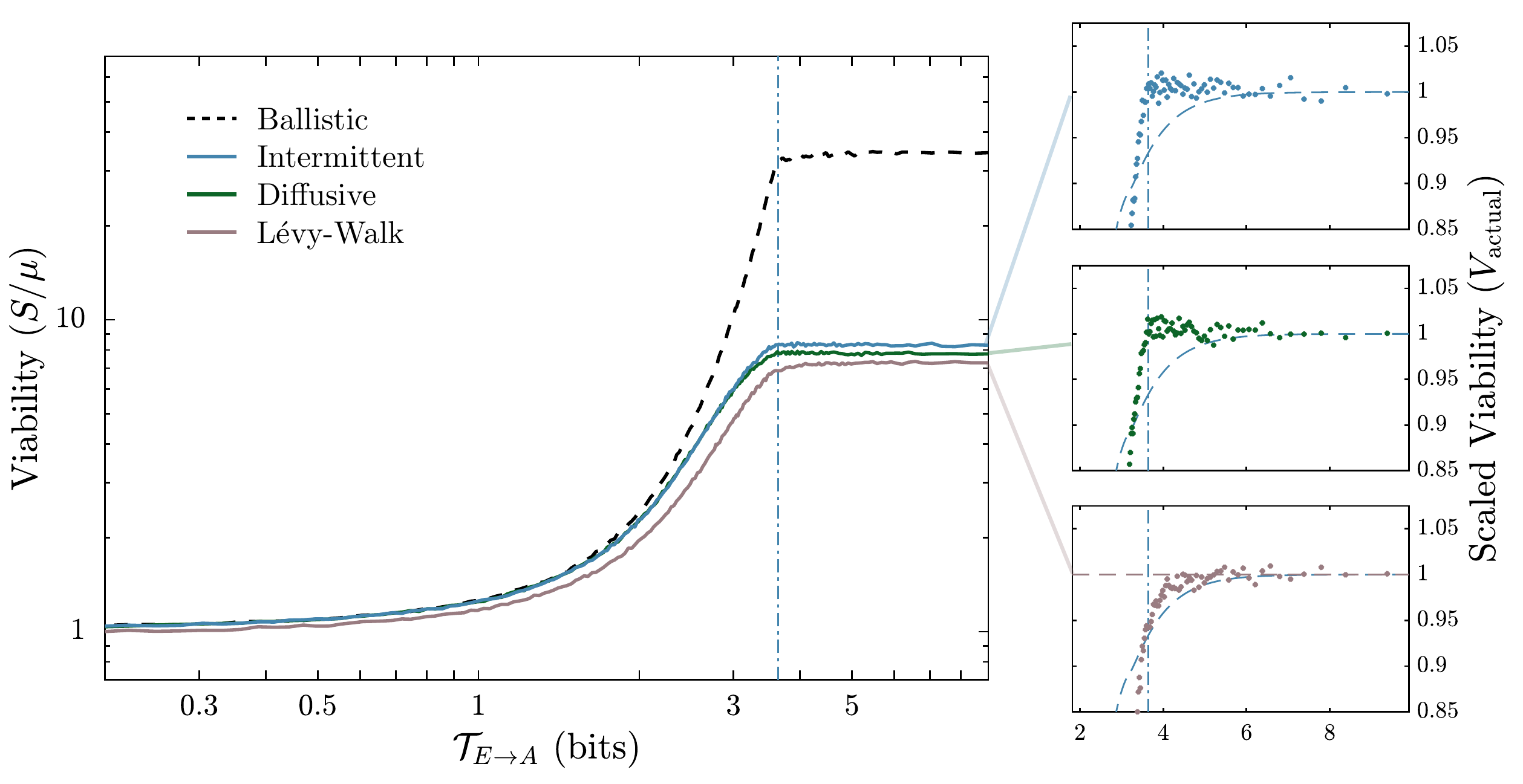}
    \caption{Comparison of the semantic threshold and viability bound for multiple foraging strategies. (Left panel) The viability curve for the ballistic forager (dashed) is presented for reference. In all strategies foragers move with a velocity $5v_{\star}$ (in the L\'evy flight $\langle v \rangle = 5v_{\star}$). (Right panels) Rescaled viability data points in the plateau region. The vertical line (dot-dashed) is the semantic threshold from Eq.~\eqref{eq: semantic threshold}, and the dashed curve is the viability bound from Eq.~\eqref{eq: broken sensor viability}.}
    \label{fig: Comparison}
\end{figure*}

\subsection{Robustness to choice of foraging strategy}

To see how robust these results are to changes in the foraging behavior, we consider several other strategies of increasing complexity. 
The most oft-studied version is that of a diffusive forager that undergoes a random walk when not detecting a resource (and then moving ballistically on detection)~\cite{benichou2014depletion, bhat2017does, benichou2016role, bhat2022smart}. 
Adding an additional layer of complexity are intermittent foragers~\cite{benichou2011intermittent, benichou2005stochastic, benichou2006two}. 
Here, the forager moves diffusively, but not always for the purpose of detecting resources. 
For large periods the movement represents locomotion, with intermittent sensing and consumption events. 
This represents the fact that biological organisms are not always looking to consume, but spend their time performing other functions. 
Finally, L\'evy-walk foragers~\cite{bartumeus2002optimizing,viswanathan2002levy} are inspired by empirical observations of animal movement~\cite{Barbosa_2018}, whereby foragers move at variable speed drawn from a (truncated) power-law distribution of velocities.   
In Appendix~\ref{app: model comparisons}, we provide details on how each of these strategies are incorporated into our simulations.

The viability curve for each strategy is plotted in the left panel of Fig.~\ref{fig: Comparison}, where the dashed line indicates ballistic, blue intermittent, green diffusive, and finally pink line L\'evy walk foraging. 
We note that the shape of the curve for all strategies is retained, and that the analytically calculated semantic threshold matches the numerical simulations for all considered strategies. 
In the right panel we show the (rescaled) plateau region for all strategies indicating good agreement with the viability bound Eq.~\eqref{eq: broken sensor viability}. 
The figure indicates that the height of the viability plateau decreases with increasingly complex strategies, with ballistic foraging leading to the highest viability and L\'evy walk the lowest. 
This is a pleasing result, given the previous observation that ballistic foraging is the most \emph{efficient} in a replenishing environment~\cite{james2008optimizing} (where efficiency is measured in terms of the encounter-rate of the forager with a resource).

Thus, the viability factorizes into a plateau that depends implicitly on both environment and agent, and a scaling term that depends only on properties of the agent. 
Furthermore, below the semantic threshold, the viability curves for multiple strategies asymptote to the same minimum viability.
This is fairly straightforward to understand. 
When locked onto a target, every implementation of our strategies defaults to ballistic motion, and maintains it until a resource is collected.
For the range of simulation parameters used, there is a high probability of a resource being found in the sensing area of a newly instantiated agent ($P_{N>0}\approx 99.2\%$) so it is most likely that the instance has $\tau=1$ and is moving ballistically.
The different strategies change what foragers do when not locked onto a target.
The formation of low-resource ``deserts'' is therefore crucial to switching between behaviors that differentiate strategies, but as sensor scrambling is increased, agents tend to travel farther each time they engage in ballistic motion, and end up in areas well outside their local desert.
Then, once again, there is a high probability that a resource is within sensing range, and the process begins anew. 
In the limit of a completely noisy sensor, all variants behave like a ballistic forager that chooses a random direction after each resource collection.

\section{Discussion and Conclusion}
\label{sec: Conclusion}
In this work, we explored the application of {a new theory of} Semantic Information to the well-motivated problem of a resource foraging agent.  
Semantic information is defined as the subset of correlations---measured here via the transfer entropy $\mathcal T_{E\rightarrow A}^\eta$---  between agent $A$ and environment $E$ that is necessary for the agent to maintain its viability $V$.  
Viability, in turn, is endogenously defined as opposed to the use of exogenous quantities like utility functions (we note a subtlety in terms of different possible definitions of ``system existence'',  which could be a topic of further exploration).  
The semantic information content in a particular agent-environment system is determined by rerunning the system evolution for intervened versions of  the dynamics. 
Such interventions involve ``scrambling'' the information transfer between the agent and environment in a specified way. 
Tracking changes in agent viability for a distribution of intervened system trajectories allows the semantic information content to be determined.  

Applying this procedure to our forager model required finding appropriate approximations to the fully specified state space of the original KW18 formalism. 
This ``operationalization'' of the KW18 prescription in terms of demonstrating how apply it to a multi-dimensional model is, itself, an important new result of our study.
For realistic systems with many degrees of freedom, the full specification of the joint probability distribution $\rho_{AE}$ over the full state space, as required in the initial formalism, may prove computationally expensive or even intractable.  
For our model we adopted a phenomenological perspective which traced over those degrees of freedom not essential to specification of the viability (this is, in part, what we mean by "operationalization").
 This left us with a reduced state space in which intervened trajectories could be simulated and the subsequent transfer entropy could be calculated.  
 To the best of our knowledge this represents the first application of semantic information in a (quasi)-realistic setting.

In our model, the forager's movements were determined by its ability to sense the presence of an individual unit of resource.  
Once detected (sensing was limited to within a radius $R$), the forager moved towards the resource.  
The transfer entropy was scrambled by adding noise to the forager's sensor via a parameter $\eta$,  where $\eta = 1$ implied a complete loss of forager's ability to sense the direction of the resource.  
Our results, expressed in terms of a viability function defined as the expectation value of the forager's lifetime, clearly showed the effect of adding noise to the sensor.  
For $\eta > \eta_c$, where $\eta_c$ is a critical value set by the sensing radius, the forager's expected lifetime was dramatically reduced (Fig.~\ref{fig: Lifetimes}). 
We refer to this as the semantic threshold. 

This result by itself represents an important extension of previous work on forager dynamics~\cite{benichou2014depletion, bhat2017does, benichou2016role, bhat2022smart, bartumeus2002optimizing, bartumeus2016foraging, benichou2005stochastic, benichou2006two, benichou2011intermittent}. 
What is novel in our work is the way in which casting the problem in terms of semantic information reveals useful aspects of the model dynamics. 
For instance, while ballistic foraging has been shown to be the most efficient strategy in terms of optimizing the encounter rate of foragers with resources (in a replenishing environment)~\cite{james2008optimizing}, here we find the same result but recast in terms of information theory and the concept of viability (or survival).
The transfer entropy represents correlations established between the forager (an agent) and the environment via the agent's sensor. 
A blind forager $\eta = 1$ has a sensor that is fully decoupled from the environment. 
By tracking how the forager's viability changes as these correlations are either increased or decreased, we gain some understanding about the role they play in the forager's ability to persist.  
In particular the upper panel of Fig.~\ref{fig: Viability} which shows the viability curve $V(\mathcal T)$, reveals two essential ways to understand the role of such correlations for the forager (or for any agent) which we now unpack separately.

Above the semantic threshold (i.e. to the right of threshold in Figure 3, 4 and 5) we find a plateau of high viability.  
Moving from right to left in this region we are removing correlations between agent and environment, however this does not effect the agent's ability to maintain its existence.  
Thus the correlations that are being removed are not essential to maintain agent viability. 
Below the semantic threshold, each bit of information  affects the agents ability to persist.  
Once this threshold is passed, we see the viability monotonically decrease. 
Thus casting the forager/environment system into the semantic information formalism allows us to see exactly how much information matters. 
In addition, our ability to model the shape of the $V(\mathcal{T})$ curve (including in particular the location of the semantic threshold) in terms of the forager/environment parameters ($R, r, \mu, \eta$) allows us to see how the correlations are instantiated.
Layering on increasingly complex foraging strategies mimicking various aspects of biological systems leads to the same qualitative results. The semantic threshold, the shape of the viability curve and the viability bound are robust to whether the forager moves ballistically, diffusively, intermittently or undertaking a L\'evy flight.
Thus, our work can provide a useful starting point for studies of more complex biological behavior in terms of semantic information in order to better understand the underlying structure of correlations between agents and their environments as well as the role information dynamics plays in their behavior.

Coming from the left in Fig.~\ref{fig: Viability} leads us to a different perspective, which may prove useful in using semantic information to understand how agents arise in the first place. 
Beginning with the low viability region on the left, we see that adding correlations  initially has little effect. 
The forager dies quickly and adding an additional bit of correlation with the environment does not change that outcome.  
As more bits of information are acquired, the agent's viability slowly rises. 
However, it is only near the semantic threshold that the slope of the curve accelerates and \emph{viability per bit} (VpB) peaks.  
Thus, it is possible that this threshold can prove useful for understanding the role information plays in allowing systems to become autonomous agents. 
Both hurricanes and cells are non-linear, driven, far-from-equilibrium systems, but only cells are considered agents. 
Future work could explore the relationship between the accumulation of semantic information and the emergence of agent-like behavior while also considering the thermodynamic cost of such accumulation.

\section*{Acknowledgements}
This project was partly made possible through the support of Grant 62417 from the John Templeton Foundation. 
The opinions expressed in this publication are those of the author(s) and do not necessarily reflect the views of the John Templeton Foundation. 
The authors thank the Center for Integrated Research Computing (CIRC) at the University of Rochester for providing computational resources and technical support.
AK thanks Sosuke Ito for support and encouragement. 
JP is supported by ``Mar\'ia de Maezt\'u'' fellowship MDM-2014-0370-17-2. 


\appendix

\section{Equilibrium resource density}
\label{sec:erd}

Consider a large patch of the environment, $A\subset\mathds{R}^2$. 
The number of resources in the patch is 
\begin{align}
    N(t)=\int_A\! \mathrm{d}^2x \ n(\bs{x},t),
\end{align}
where $n$ is the resource density.
Resource do not move, but they can be created or destroyed.
Creation is due to an energy flux impinging on $A$, denoted $\Gamma(\bs{x},t)$; there exists some mechanism in the environment that converts this flux into localized resource deposits of energy $\epsilon$.
Destruction is due to another mechanism by which resources decay, their energy lost as heat, denoted $\gamma(\bs{x},t)$.
Note that the former rate is independent of the number of resources, the latter is not.
The rate of change of resources in this large patch is
\begin{align}
    \frac{dN}{dt} = \int_A\! \mathrm{d}^2x\ \left(\frac{\Gamma(\bs{x},t)}{\epsilon}-\gamma(\bs{x},t)n(\bs{x},t)\right).
\end{align}
Taking $\Gamma$ and $\gamma$ both to be static, the resource density satisfies 
\begin{align}
    \frac{\partial n}{\partial t} &=  \frac{\Gamma }{\epsilon}-\gamma n\nonumber\\
    &\Downarrow\nonumber\\ n(t) &= n_0 e^{-\gamma t} + n_{eq}(1-e^{-\gamma t}).
\end{align}
where the equilibrium density is $n_{eq}=\Gamma/\epsilon\gamma$.
The average area occupied by a single resource is the reciprocal of this, $\epsilon\gamma/\Gamma$, so that the average spacing between resources at equilibrium is
\begin{align}
    \ell_{Re} = \sqrt{\frac{\epsilon\gamma}{\Gamma}}.
\end{align}

If instead we coarse grain over the positions of resources and care only about their count, then the transition probabilities for a resource being generated or destroyed in a time $\Delta t\rightarrow 0$ are $p(N\rightarrow N+1) = \Gamma A\Delta t / \epsilon$ and $p(N\rightarrow N-1) = \gamma N\Delta t$.
The transitions between these coarse grained states define an $M/M/\infty$ queue --- a well known stochastic process with a stationary distribution that is Poisson: 
\begin{align}
    p(N) = \frac{1}{N!}\left(\frac{\Gamma A}{\epsilon\gamma}\right)^Ne^{-\frac{\Gamma A}{\epsilon\gamma}}.
\end{align}
The expected number of resources is $\langle N\rangle =\frac{\Gamma A}{\epsilon\gamma}$, which under the assumption of a homogeneous distribution, gives an equilibrium number density that matches $n_{eq}$ from the preceding paragraphs.
The variance in the number of resources is $\delta N^2=n_{eq} A$, giving a fluctuation in the number density of $\delta n=\ell_{Re}^{\mhyphen 1}A^{-1/2}$.
Thus when considering large environmnets, $A\gg\ell_{Re}^2$, the relative fluctuations around the equilibrium density are negligible, $\delta n/n\rightarrow 0$.

\section{Simulation details}
\label{app:sims}
Translating our model for numerical simulation requires first dimensionalizing all parameters in Table~\ref{table: model parameters}. 
Natural scales were chosen to be agent-centric, so that numerical values within each simulation are all interpreted in terms of scales meaningful to the forager.
The natural timescale is T$=S/\mu$, the lifetime of an agent on a full tank---with no other influxes of fuel; the natural length scale is L $=r$, the extraction radius of the agent; and the natural mass scale, M$= S$ T$^2$L$^{\mhyphen 2}$, the size of a full fuel tank.

Dimensionalization is accomplished by dividing model parameters by the combination of natural units given in the far right column of Table~\ref{table: model parameters}.
For example, the dimensionless (tilded) speed is $\tilde v = v/($LT$^{\mhyphen 1})$, read off naturally as the number of extraction radii ($\sim$ forager length) travelled on a full tank of fuel.
We summarize the chosen scales and dimensionless values of parameters, as well as derived dimensionless values of relevant quantities in Table~\ref{table: simulation parameters}.
\begin{table*}[ht]
\caption{Primary Simulation Dimensionless Parameters}
\centering 
\begin{tabular}{l|lcc}    \toprule
& \emph{Parameter} 		& \emph{Symbol} 	& \emph{Value} \\\midrule
&Collection Radius    		& $\tilde{r}$  		& $1$  \\
\emph{Scaling}&Metabolic Rate       		& $\tilde{\mu}$ 		& $1$ \\
&Maximum Stored Energy	& $\tilde{S}$ 		& $1$ \\ \midrule
&Detection Radius      		& $\tilde{R}$  		& $4$  \\
&Speed                		& $\tilde{v}$ 		& $20$(Fig.\ref{fig: Viability}) $,25$ (Fig.\ref{fig: Comparison}) \\
&Time Discretization		& $\Delta\tilde{t}$	& $0.02$\\ 
\emph{Chosen}&Environment Size 		& $\tilde{L}$		& $100$ \\
&Resource Energy      	& $\tilde{\epsilon}$ 	& $0.1$ \\
&Resource Decay Rate  	& $\tilde{\gamma}$	& $0.1$ \\
&Energy Influx        		& $\tilde{\Gamma}$	& $0.001$ \\ \midrule
&Minimal Speed			& $\tilde{v}_\star$	& $5$ \\
\emph{Derived}&Maximal Speed	 		& $\tilde{v}^\star$	& $50$ \\
&Equilibrium Resource Density & $\tilde{n}_{eq}$ & $0.1$\\
&Equilibrium Resource Spacing& $\tilde{\ell}_{Re}$ & $\sqrt{10}\approx3.16$\\
\bottomrule
\hline
\end{tabular}\label{table: simulation parameters}
\end{table*}

After dimensionalizing, data is generated by simulating $10^4$ instances of a single forager placed into an environment at equilibrium, for multiple values of the scrambling parameter $\eta$.
Specifically, $200$ values of $\eta$ were chosen, with the first $100$ spaced linearly on the interval $[0,0.15)$, and the second $100$ logarithmically space on the interval $[0.15, 1]$.
These were chosen after initial coarser simulations revealed the location of the semantic threshold so as to ensure the resolution of the viability plateau.
For each value of $\eta$, each of the ten thousand simulations are begun by instantiating a forager at half health, $\tilde{s}=0.5$, at the center of an environment --- a square with side length $\tilde{L}=100$--- with $N=\tilde{n}_{eq}\tilde{L}^2=10^3$ resources distributed uniformly.

Each individual simulation evolves with a temporal discretization $\Delta\tilde{t}=1/50=0.02$ following the flowchart presented in Fig.~\ref{fig:forager_model}.
Below we summarize the chart, describing what our code does in a single timestep for both the degrees of freedom (DoFs) of the agent, $(s,\tau,\hat{\bs{n}},\bs{x})$, and environment, $(\bs{y}_1,\bs{y}_2,\dots)$ (whose DoFs are ordered by distance from the agent).
First, the agent must be alive, $s>0$, otherwise the simulation ends.
Otherwise the closest resource, $\bs{y}_1$ is examined.
The probability that two or more resources have the same closest distance is for all intents and purposes $0$, but should this miracle occur only one amongst these is chosen, uniformly at random.
If that resource is within the extraction radius of the agent, then it is harvested, whereupon the agent's fuel supply increases $s\mapsto s+\epsilon$, the agent's target DoF is set to $\tau\rightarrow 0$, while its orientation and position remain unchanged, $(\hat{\bs{n}},\bs{x})\mapsto(\hat{\bs{n}},\bs{x})$.
If that resource is not within extraction range, then the agent's target DoF, $\tau$, is examined.
If a target is set, $\tau=1$, then it remains set, as does the orientation of the agent; simultaneously the fuel supply decreases, $s\mapsto s-\mu\Delta t$, as the agent moves, $\bs{x}\mapsto\bs{x}+v\hat{\bs{n}}\Delta t$.
If the target is not set, $\tau=0$, then the code checks if the closest resource is within sensing range of the agent.
If it is, then the agent sets a target $\tau\mapsto 1$, and expends fuel $s\mapsto s-\mu\Delta t$, to reorient and move towards it, $(\hat{\bs{n}},\bs{x})\mapsto(\bs{y}_1/|\bs{y}_1|,\bs{x}+\bs{y}_1/|\bs{y}_1|\Delta t)$.
If the closest resource is beyond sensing range, then the target remains unset, the agent's orientation does not change as it expends fuel, $s\mapsto s-\mu\Delta t$, to move in that direction, $\bs{x}\mapsto\bs{x}+\hat{\bs{n}}v\Delta t$.

Since we have a finite environment, sometimes the agent may find itself approaching a boundary, in which case the code implements reflective boundary conditions as follows.
Should one of the components of the position of the agent fall outside the interval $[0,L]$, then it is replaced by the reflected position in the interval. That is,  if $x<0$ then $x\mapsto-x$, while if $x>L$ then $x\mapsto L-x$.
If a spatial reflection takes place, then the corresponding component of the velocity is also reflected, $v_x\mapsto-v_x$.
This completes the agent's update timestep.

Next, the environment updates via a full timestep consisting of a degradation step followed by a growth step.
During the degradation step, each resource is checked to see if it degrades.
A check consists of drawing a uniform random number and seeing if it greater than $e^{-\gamma\Delta t}$.
If it is, the resource is removed from the environment, otherwise it remains.
During the growth step, a random non-negative integer is chosen from a Poisson distribution with $\lambda=\Gamma L^2\Delta t/\epsilon=2$, i.e.,
\begin{equation}
	p_n = \frac{e^{-\lambda}}{n!}\lambda^n.
\end{equation}
This integer represents how many new resources are generated; these are added to the environment at uniformly random locations. 
The displacement and distance from the agent to each resource is updated, and the list of resource locations is reordered according to proximity.
This completes the environment update timestep.

For each of the $10^4$ simulations at each value of $\eta$, the degrees of freedom of the agent and the location of the nearest resource are saved at each time step. 
On the plateau, simulations run for $\sim 40$ minutes, averaging $\sim 750$ timesteps and resulting in datasets of $\sim0.5$Gb; simulations on the other side of the plateau dropped in both temporal and spatial complexity.
For a fixed set of parameters, $10^4$ simulations across the $200$ scrambling parameters requires $\sim8000$ core-hours of computation and generates $\sim60$Gb of data.
Simulations were conducted on the Center for Integrated Research Computing's BlueHive cluster at the University of Rochester. 

\section{The Semantic Threshold}
\label{sec:ST}
\begin{figure}

\centering
    
\begin{center}
\begin{tikzpicture}

\coordinate (Agent) at (0,0);
\coordinate (Resource) at (1.2,2.7495);
\coordinate (r) at (1.2,0);
\coordinate (Target) at (0,4);

\node[rectangle,
    bottom color=ForagerPink2,
    top color=white,
    minimum width = 2.4cm, 
    minimum height = 4cm] (harvest) at (0,2) {};

\draw[thin] (Agent) circle (3.0);
\draw[fill=white, thin] (Agent) circle (1.2);

\draw[very thick] (Agent) -- (r) node[midway,below]{$r$};
\draw[very thick] (Agent) -- (1.2,2.7495) node[midway,left]{ $R$} -- (1.2,0) ;
\draw[very thick] (r) -- (1.2,2.7495) node[midway,left]{} ;
\draw[dashed] (Agent) -- (Target) node[midway,left]{};

\filldraw [SkyBlue] (Agent) circle (3pt) node[below left] {Agent};
\filldraw [ForagerPink] (0,3) circle (3pt) node[above left] {Target};
\filldraw [ResourceGreen] (Resource) circle (3pt) node[above right] {Resource};

\pic [thick, draw=black,double, angle radius = 0.7cm] {angle = Resource--Agent--Target};
\pic [thick, draw=black,double, angle radius = 0.7cm] {angle = Agent--Resource--r};
\pic [thick, draw=black, angle radius = 0.3cm] {right angle = Resource--r--Agent};
\end{tikzpicture}
\end{center}
\caption{The geometry of an agent (mis)targeting a resource at the edge of sensor range.}
\label{fig: target geometry 1}
\end{figure}
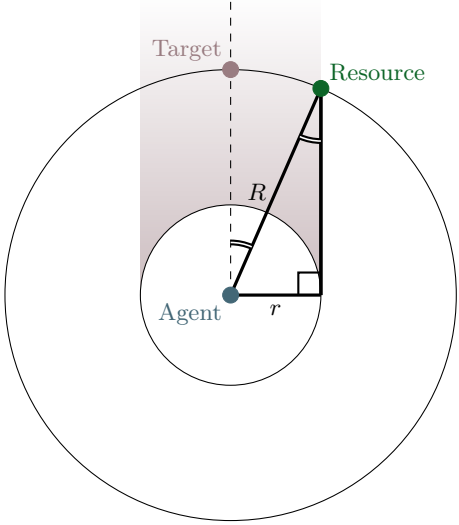

Consider an agent for which a resource has just entered sensor range and a targeting event has occurred.
We scramble the information relayed by the event so that the agent reorients itself with a target that is misaligned with the resource.
The agent begins its journey towards the target, as shown in Fig.~\ref{fig: target geometry 1}.
For an agent moving towards a target any resource that falls within the grey region will be collected.
The maximal misalignment angle between target and resource is $\pi\eta$.
From the diagram once can infer a critical level of scrambling, $\eta_c$, satisfying
\begin{equation}\label{eq: critical angle}
    \sin\pi\eta_c = \frac{r}{R},
\end{equation}
below which the agent will always collect the resource.
Above it, the agent will sometimes miss the resource --- an event that could potentially result in starvation.
Combining Eq.~\eqref{eq: critical angle} with Eq.~\eqref{eq: mutual information} gives the semantic threshold, Eq.~\eqref{eq: semantic threshold}.

\section{The Viability Plateau}
\label{sec:vp}

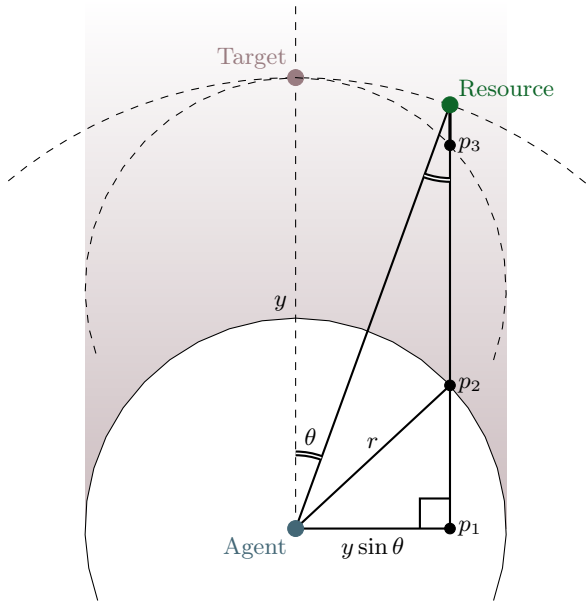
\begin{figure}
\begin{center}
\begin{tikzpicture}
\coordinate (Agent) at (0,0);
\coordinate (Resource) at (2.0521,5.6382);
\coordinate (r) at (2.8,0);
\coordinate (Target) at (0,6);
\coordinate (Rint) at (2.0521,1.905);
\coordinate (p1) at (2.0521,0);
\coordinate (p2) at (2.0521,5.1);

\node[rectangle,
    bottom color=ForagerPink2, top color=white,
    minimum width = 5.6cm, 
    minimum height = 7cm] (harvest) at (0,3.5) {};
\draw [black,thin,domain=-20:200,fill=white] plot ({2.8*cos(\x)},{2.8*sin(\x)});
\draw [black,thin,dashed,domain=50:130] plot ({6.0*cos(\x)},{6.0*sin(\x)});

\draw[dashed] (Agent) -- (Target) node[midway,left]{$y$};
\draw[dashed] (Target) -- (0,7);
\draw[thick] (Agent) -- (Resource) node[midway,left]{};
\draw[thick] (Agent) -- (2.0521,0) node[midway,below]{$y\sin\theta$};
\draw[thick] (Resource) -- (2.0521,0) node[midway,left]{};
\draw[thick] (Agent) -- (Rint) node[midway,above]{$r$};
\draw[very thick] (Resource) -- (p2);

\filldraw [SkyBlue] (Agent) circle (3pt) node[below left] {Agent};
\filldraw [ForagerPink] (0,6) circle (3pt) node[above left] {Target};
\filldraw [ResourceGreen] (Resource) circle (3pt) node[above right] {Resource};

\filldraw [black] (p1) circle (2pt) node[right] {$p_1$};
\filldraw [black] (Rint) circle (2pt) node[right] {$p_2$};
\filldraw [black] (p2) circle (2pt) node[right] {$p_3$};

\pic [thick, draw=black,double, angle radius = 1.0cm] {angle = Resource--Agent--Target};
\pic [thick, draw=black,double, angle radius = 1.0cm] {angle = Agent--Resource--Rint};
\pic [thick, draw=black, angle radius = 0.4cm] {right angle = Resource--p1--Agent};
\draw (0.2,1) node[left, above]{$\theta$};

\draw [black,thin,domain=-20:200,dashed] plot ({2.8*cos(\x)},{3.2+2.8*sin(\x)});
\end{tikzpicture}
\end{center}
    \caption{The geometry describing how far an agent has to travel to collect a mistargeted resource.}
    \label{fig: target geometry 2}
\end{figure}

Let's examine a case below the critical scrambling strength, so that every targetted resource is still collected. 
Once again we consider a scrambled targeting event, but assume it occurs for a resource a distance $y<R$ away from the agent.
This type of event occurs regularly after the agent has collected a resource and there is at least one other resource in sensor range.
The agent begins to move towards the target as depicted in Fig.~\ref{fig: target geometry 2}.

Had the sensor not been scrambled, the agent would traverse a distance $y-r$ to collect the resource.
Meanwhile, the scrambled agent needs to travel the distance from $p_2$ to the resource for collection to occur; a little geometry and trigonometry on the diagram give us this distance as $y\cos\theta - \sqrt{r^2-y^2\sin^2\theta}$.
Since $p_2$ to $p_3$ is also a distance $y-r$, denote the remaining distance $\lambda(\theta) y$ where $\lambda(\theta)$ is referred to as a {\it dilation} factor.
It will become clear why in what follows.

A little algebra gets us the angular dependence of the dilation factor,
\begin{align}
    \lambda(\theta) = &\frac{r}{y}-1+\cos\theta-\sqrt{\frac{r^2}{y^2}-\sin^2\theta}.
\end{align}
Of course this factor will be different each time a targeting event occurs, depending on distance to resource and the particular mismatch angle of the target.
However, for a given $\eta$ we know the distribution of mismatch angles --- it is uniform over $[-\pi\eta,\pi\eta]$ --- so for fixed $y$, the expected dilation is
\begin{align*}
    \mathds{E}_\eta[\lambda] = \frac{1}{2\pi\eta}\int_{\mhyphen\pi\eta}^{\pi\eta}\!\!\! d\theta \ \lambda(\theta).
\end{align*}
Examining the functional form of $\lambda(\theta)$ we see that the first two terms are trivial and the third term easily integrates to $\sin\pi\eta/\pi\eta$, while the fourth term requires a little massaging.
After factoring out a $-r/y\pi\eta$ and using that the  integrand is even,  the remaining factor is massaged into the form
\begin{align*}
    \int_{0}^{\pi\eta}\!\!\! d\theta \ \sqrt{1-\frac{y^2}{r^2}\sin^2\theta}.
\end{align*}
The integral is not elementary, but it turns out to be an incomplete elliptic integral of the second kind,
\begin{align}
    E(\varphi|k^2) = \int_0^\varphi\!\!\! d\theta\ \sqrt{1-k^2\sin^2\theta},
\end{align}
with modulus $k$.
There's a slight nuance since normally $0<k^2<1$, while in our case $y>r$ gives a modulus greater than unity.
This bound ensures the integrand remains real for all values of $\varphi$.
Fortunately, since we're considering cases below critical scrambling, $y\sin\theta<r$ is automatically satisfied and there's no need to worry about which branch we're on.
With this minutiae out of the way, we write the expected dilation for fixed $y$
\begin{align}\label{eq: app lambda}
    \mathds{E}_\eta[\lambda] = \frac{r}{y}-1+\frac{\sin\pi\eta}{\pi\eta}-\frac{r}{y}\frac{E(\pi\eta|\frac{y^2}{r^2})}{\pi\eta},
\end{align}
which when combined with Eq.~\eqref{eq: mutual information} and \eqref{eq: semantic threshold}, yields Eq.~\eqref{eq: dilation0}.
To simplify the notation, we simply denote this as $\lambda$ in the main text.

The asymptotic behaviors of the dilation require Taylor expanding the elliptic integral.
For the perfect sensor limit we treat $\eta\ll1$, so that
\begin{align*}
    E(\pi\eta|k^2)
    &\approx\int_0^{\pi\eta}1-\frac{k^2}{2}\theta^2\\
    &=\pi\eta-\frac{k^2}{6}\pi^3\eta^3
\end{align*}
to third order in $\eta$.
Doing the same to the sine function, and plugging into Eq.~\eqref{eq: app lambda}, one finds the first non-vanishing term is second order in $\eta$:
\begin{align*}
\frac{\pi^2}{6}\frac{y-r}{r}\eta^2.
\end{align*}
Plugging in the inverse of Eq.~\eqref{eq: mutual information} gives the quoted exponential asymptotic behavior.

The asymptotic when $y\rightarrow r^+$ requires a bit more work.
We parameterize the approach as $y=(1+\varepsilon)r$ with $\varepsilon\ll1$, and expand the elliptic integral to first order
\begin{align*}
E(\pi\eta|(1+\varepsilon)^2)&=
    \int_0^{\pi\eta}\!\!\!\!\!d\theta\ \sqrt{1\!-\!(1\!+\!\varepsilon)^2\sin^2\theta}\\
    &\approx \int_0^{\pi\eta}\!\!\!\!\!d\theta\ \sqrt{\cos^2\theta\!-\!2\varepsilon\sin^2\theta}\\
    &\approx \int_0^{\pi\eta}\!\!\!\!\!d\theta\ \cos\theta\!-\!\varepsilon \int_0^{\pi\eta}\!\!\!\!\!d\theta \ \cos\theta\tan^2\theta\\
    &=(1-\varepsilon)\sin\pi\eta-\varepsilon\ln|\sec\theta+\tan\theta|
\end{align*}
After plugging this in, Eq.~\eqref{eq: app lambda} simplifies to 
\begin{align*}
    \varepsilon\left(\frac{1}{\pi\eta}\ln\frac{\cos\pi\eta}{1+\sin\pi\eta}-1\right)
\end{align*}
The argument of the logarithm can be massaged using elementary trigonometric identities into $(1-\tan\frac{\pi\eta}{2})/(1+\tan\frac{\pi\eta}{2})$.
Then recalling that $\tanh^{\mhyphen1}x$ is equivalent to $\frac{1}{2}\ln\frac{1-x}{1+x}$, we have the quoted linear asymptotic behavior.

\section{Model Comparisons}
\label{app: model comparisons}
To examine the effect of changing the complexity of the agent on the viability curve, we performed the entire set of simulations described in the last appendix on three other forager models of increasing complexity.
Whereas the ballistic forager of the primary simulations is a completely deterministic automaton, the three models detailed below introduce internal stochasticity into the agent behavior.

\subsection*{Diffusive Forager}
\begin{figure}[ht]
    \includegraphics[width = 1.0\columnwidth]{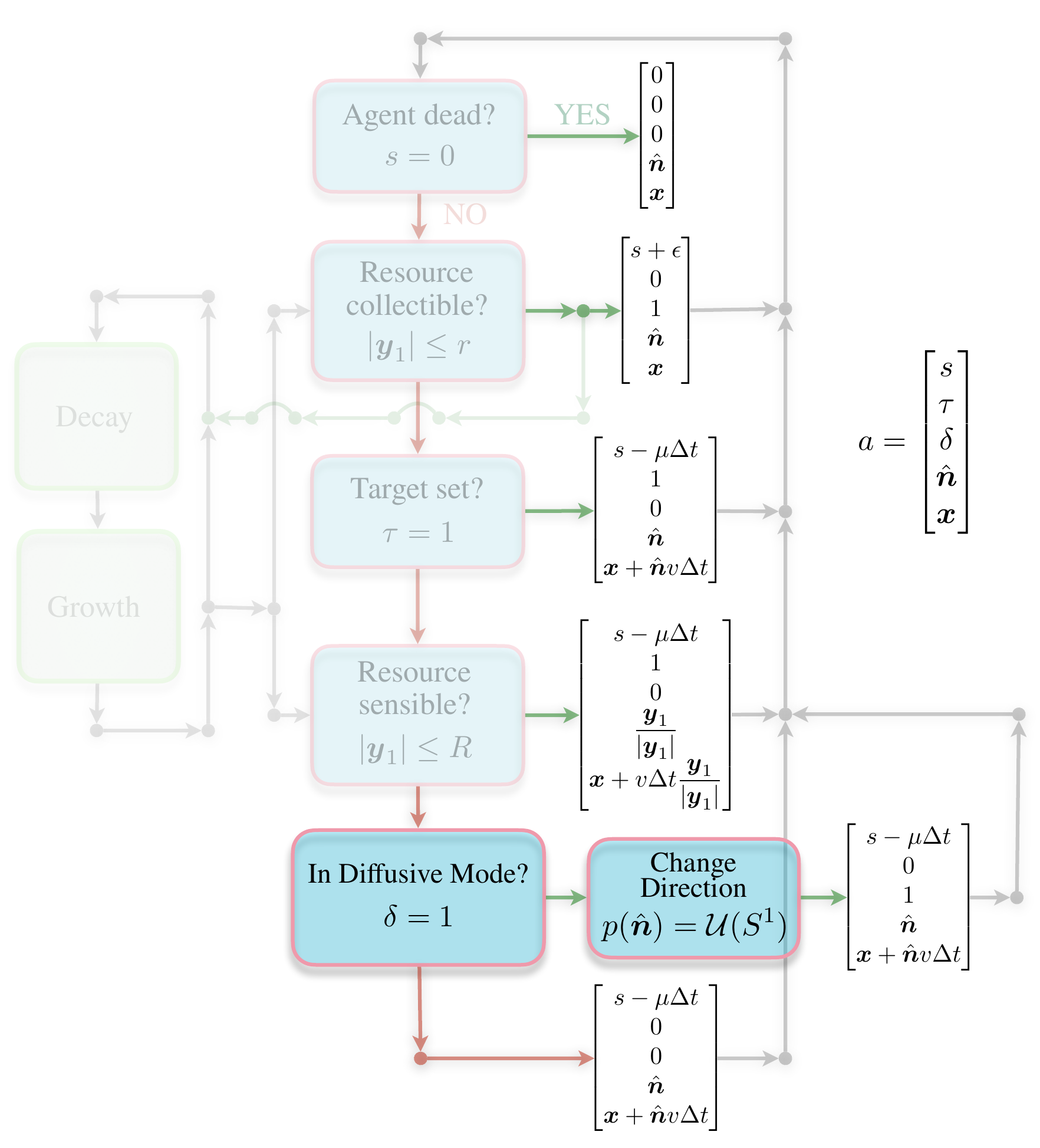}
    \caption{Diffusive Foraging Strategy}
    \label{appfig: DF}
\end{figure}

The first model increases the DoFs of the agent to include a logical switch, $\delta\in\{0,1\}$, which determines the type of locomotive behavior.
When $\delta=0$ the agent behaves as the ballistic forager in the primary simulations; when flipped to $\delta=1$ the agent performs a continuous random walk --- the namesake of the model.
The new DoF is tied to the targeting DoF, $\delta\mapsto 1-\tau$ in each agent state update of Fig.~\ref{fig:forager_model}, so that diffusive motion occurs when the agent does not have a target. 
When $\tau=0$, the orientation of the agent becomes a random variable --- an angle drawn from the uniform distribution over the circle, $\hat{\bs{n}}\sim\mathcal U(S^1)$.
There is an interesting question of what effect the scrambling of the internal correlation between $\tau$ and $\hat{\bs{n}}$ would would have on viability.
We do not explore it in this work, but mention it as an entry point for the exploration of viability-based behavior~\cite{egbert2023behaviour}. 
The architecture of the diffusive forager used in our code is shown in Fig.~\ref{appfig: DF}.

\subsection*{Intermittent Forager}
\begin{figure}[ht]
    \includegraphics[width = 1.0\columnwidth]{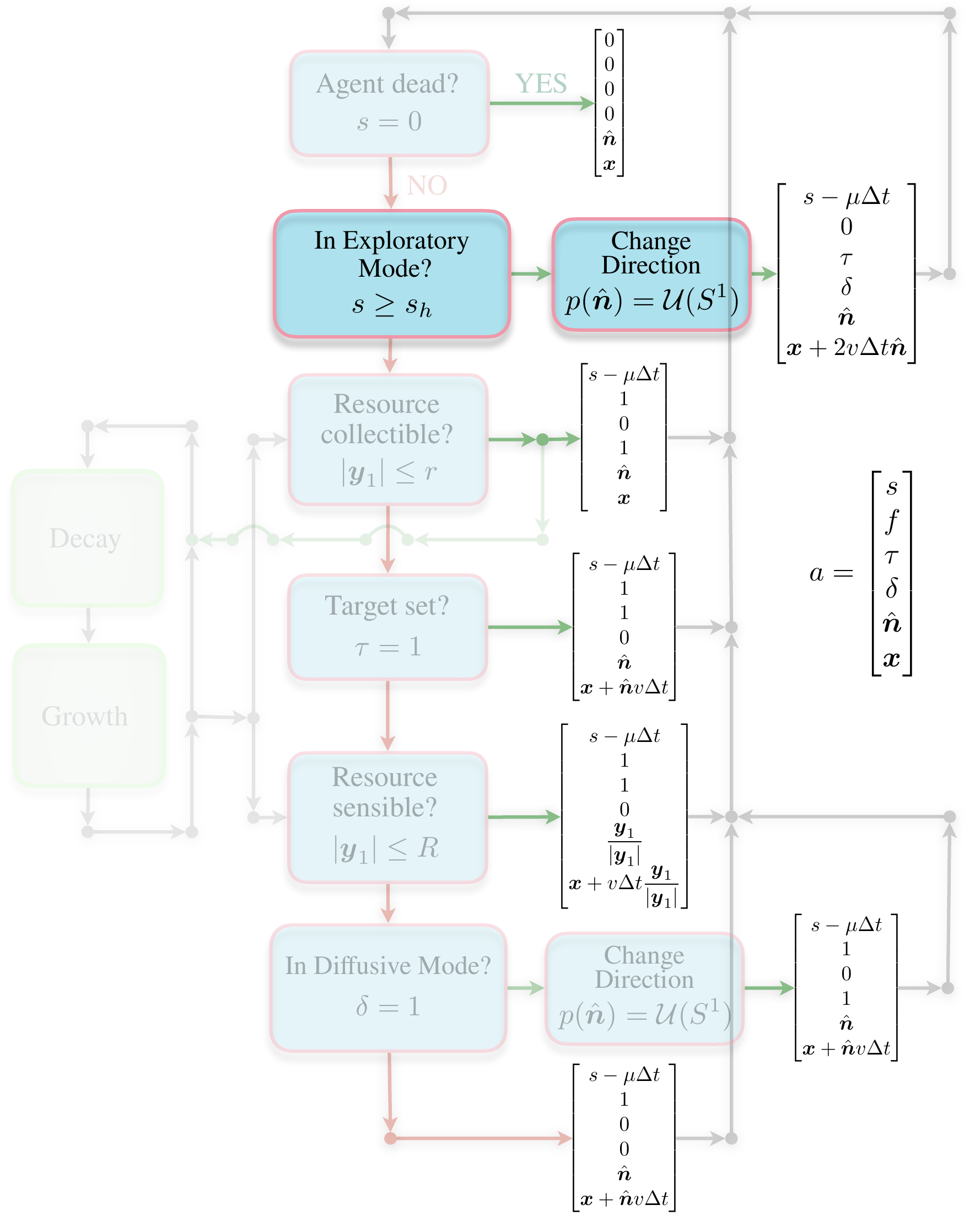}
    \caption{Intermittent Foraging Strategy}
    \label{appfig: IF}
\end{figure}

The next model increases the complexity of the diffusive forager to include another logical switch, $f\in\{0,1\}$, determining the behavioral modality of the agent.
This model is meant to imitate well-known intermittent foraging strategies.
When $f=1$, the agent behaves exactly like the diffusive forager in the prior subsection.
This modality of behavior occurs when the fuel supply is below a {\it hunger threshold}, $s\le s_h$, and is called {\it foraging}; when fuel supply of the forager is above  threshold, the modality switches to {\it exploring}.
In the exploring modality, the direction of motion is chosen randomly and independently of the environment, and the energy supplied by metabolism is no longer split between sensing and locomotion.
To incorporate this rerouting of resources, the speed of the agent is doubled, turning an exploring forager into a diffusive forager with twice the diffusivity.
Our implementation is shown in Fig.~\ref{appfig: IF}.

We experimented with different values of the hunger threshold.
The time spent in the exploring modality depends on the resource energy $\epsilon$, with the maximum time spent exploring equal to $t_E = \epsilon/\mu$, or $0.1$ full tank lifetimes for our chosen parameters.
These exploring excursions would typically happen towards the beginning of the foragers lifetime.
In order to compare the tails of the strategies, we need to account of the fact that our intermittent foragers spent a small fraction of their lives exploring. For this reason, the plateau of the reported viability curve falls just barely on top of a purely diffusive strategy.
In experiments with larger values of $\epsilon$ and higher speeds, a greater distinction appeared between the curves, with the plateau of the intermittent strategy lying between the diffusive and ballistic strategies.

\subsection*{L\'evy-walk Forager}
The final model goes back to the ballistic forager and elevates the speed parameter to a dynamical degree of freedom.
The speed is a stochastic multiple of $v_0$, drawn from a power law with exponent $\nu$, namely
\begin{equation}
P(k;k_{\min{},}k_{\max{}},\nu)=\frac{k^{-\nu}}{\sum_{n=k_{\min{}}}^{k_{\max{}}}n^{-\nu}},
\end{equation}
meant to simulate a random L\'evy walk.
This novelty is added at the end of the Ballistic Forager timestep, updating the velocity to some multiple, $k_{\min{}}\le k\le k_{\max{}}$.
\begin{figure}[ht]
    \includegraphics[width = 1.0\columnwidth]{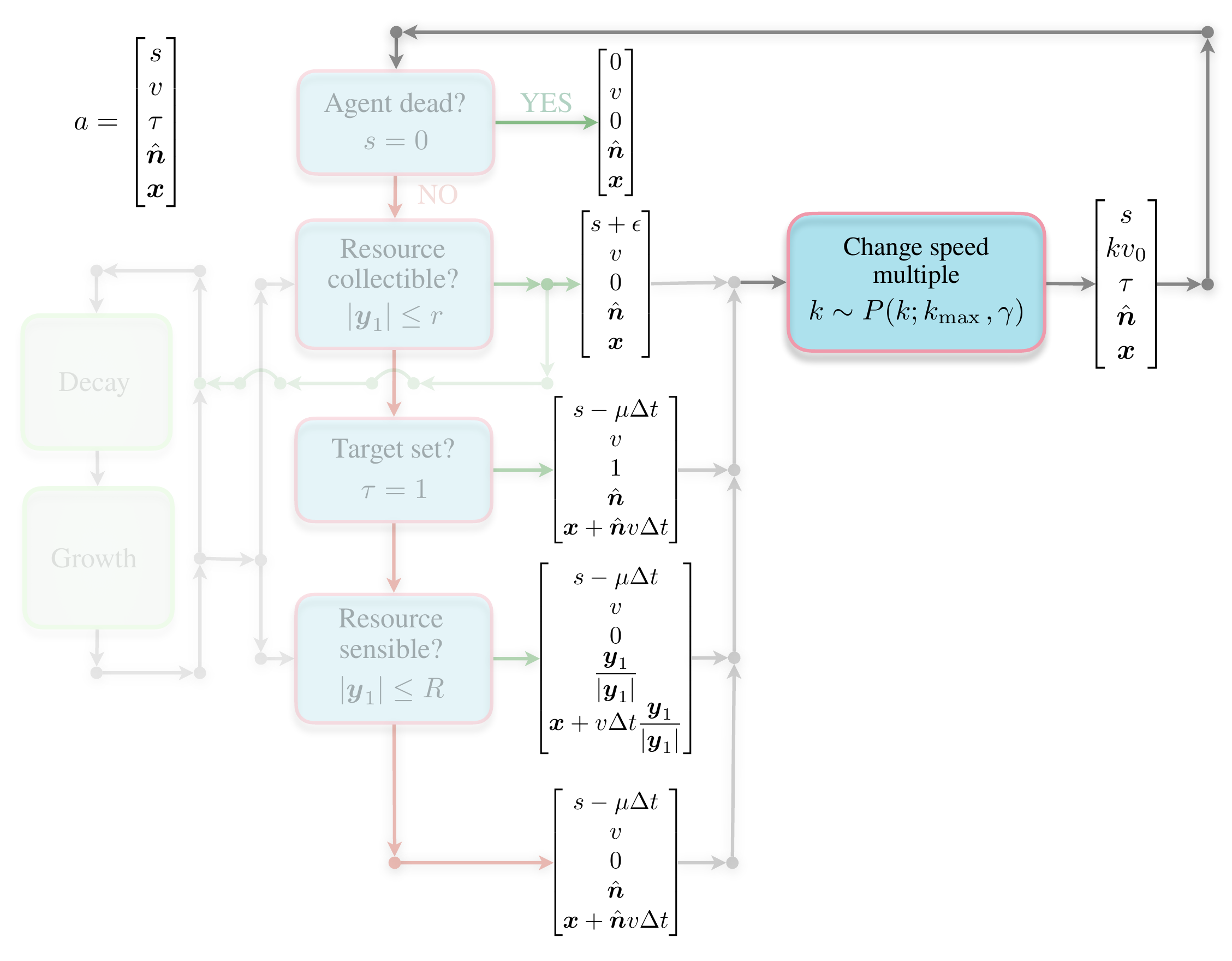}
    \caption{L\'evy-Walk Foraging Strategy}
    \label{appfig: LWF}
\end{figure}

For our experiments, we wanted to compare this strategy to the others in the limit of a completely noisy sensor.
Initial experiments revealed that this required the average velocity of all the models to be the same in order for the viabilities to match in that limit (The precise reason is elucidated in the text).
To this end we chose $\nu = 1.9658$, with $k_{\min{}}=2$, $k_{\max{}}=30$, meaning that the maximal velocity achieved by the agent was six times $v^\star$.
These ensured that $\langle v\rangle/v_\star=5.0$, the same as the other simulations, to four decimal places of accuracy.
Our implementation is shown in Fig.~\ref{appfig: LWF}.

\end{document}